\documentclass[12pt]{article}

\newcommand{\m}{\mathrm}
\newcommand{\be}{\begin{equation}}
\newcommand{\ee}{\end{equation}}
\newcommand{\ba}{\begin{eqnarray}}
\newcommand{\ea}{\end{eqnarray}}

\usepackage{graphicx}
\usepackage{amssymb}
\usepackage{amsmath}
\usepackage[T1]{fontenc} 
\usepackage[ansinew]{inputenc} 
\usepackage[nosort]{cite}
\newcommand{\inbar}{\vrule height1.57ex width.4pt depth0pt}
\newcommand{\SW}{\relax{\hbox{$\ \inbar\kern-.285em{\rm S}$}}}

\begin{document}
\thispagestyle{empty}
\begin{center}

\null \vskip-1truecm \vskip2truecm

{\Large{\bf \textsf{The Weak Gravity Conjecture Requires the Existence of Exotic AdS Black Holes}}}

{\large{\bf \textsf{}}}

{\large{\bf \textsf{}}}

\vskip1truecm

{\large \textsf{Brett McInnes}}

\vskip1truecm

\textsf{\\  National
  University of Singapore}

\textsf{email: matmcinn@nus.edu.sg}\\

\end{center}
\vskip1truecm \centerline{\textsf{ABSTRACT}} \baselineskip=15pt
\medskip

The Weak Gravity Conjecture arises from the requirement that it be possible for all (classically stable) extremal black holes to decay. The ``black hole version'' of the conjecture requires that it should be possible for this to occur through the emission of smaller black holes. We consider this version in the case of extremal AdS$_4$-Kerr-Newman black holes which are stable against a superradiant instability. One finds that the emitted black hole must be rather exotic, having an ``angular horizon'' analogous to the more familiar (radial) horizon.

\newpage

\addtocounter{section}{1}
\section* {\large{\textsf{1. Rotation and the Weak Gravity Conjecture}}}
Quantum-corrected black holes are currently the subject of intense study, from a wide range of perspectives. They are of interest not only in investigations of quantum gravity itself (for example, in the Swampland Programme \cite{kn:swamp}) but also, through gauge-gravity duality\cite{kn:casa,kn:nat,kn:bag}, in applications: for example, they appear in recent exciting developments in the study of many-body chaos \cite{kn:kob1,kn:kob2}. This particular application underlines the importance of the \emph{asymptotically AdS} case, which will be our focus here.

The most well-developed approach to the study of these objects makes use of the ``Weak Gravity Conjecture'' \cite{kn:motl,kn:palti,kn:rude,kn:rem,kn:shiu,kn:goon} (henceforth, WGC\footnote{For a sample of previous work on the asymptotically AdS version of the WGC, see \cite{kn:qing,kn:naka1,kn:mig,kn:crem,kn:agar,kn:naka2,kn:102}.}), which is closely related to the assertion that (otherwise apparently stable) extremal black holes must in fact be able to decay.

The black hole decay products required by the WGC could be elementary particles, but, in the ``black hole version'' of the WGC, they might also themselves be black holes. For electrically or magnetically charged black holes, the daughter black holes necessarily violate classical Cosmic Censorship. This version of the WGC therefore implies \cite{kn:kats} a particular form for quantum corrections to classical Censorship conditions\footnote{There are good reasons to think that, whatever may be the case for asymptotically flat spacetimes, some form of Cosmic Censorship does hold for the asymptotically AdS spacetimes of interest to us here: see for example \cite{kn:weak,kn:horsant,kn:suvrat,kn:bala,kn:sean,kn:wang,kn:netta}.}: the WGC-modified version of extremality is such that, for a given mass, larger values of the charge are allowed than in the classical case\footnote{Alternatively, one can try to give explicit expressions for extremality when $\alpha^{\prime}$ corrections to the black hole geometry are taken into account \cite{kn:cano1,kn:cano2}. Unfortunately this is currently possible only under rather restrictive conditions, not in the cases with which we are concerned here.}.

It is natural to ask whether there are analogous results for other kinds of charge, and indeed there is evidence that this is the case for topological charges \cite{kn:crem2}. It is still more natural to ask whether something similar occurs when extremality is due to very high \emph{angular momenta}, instead of (or in addition to) charge. This has been investigated in the case of the BTZ spacetime \cite{kn:jorge,kn:roberto} in \cite{kn:aalsma}, but little is known in the four-dimensional case.

The reason for this is that four-dimensional AdS-Kerr extremal black holes have a peculiar property (not shared by BTZ black holes \cite{kn:ortiz}): they are already classically \emph{unstable}, due to superradiance \cite{kn:hawkreall,kn:super,kn:amado}. Thus, the instability mandated by the WGC in the asymptotically flat case does not have an opportunity to arise here.

Recently, however, it was discovered \cite{kn:103} that superradiance can be avoided if the black hole is both rotating and (electrically or magnetically) charged: that is, in the extremal AdS$_4$-\emph{Kerr-Newman} case. This occurs provided that the charge is sufficiently large. For these (otherwise stable) black holes, then, we can again ask the question: granted that classically extremal AdS$_4$-Kerr-Newman black holes must be able to decay, what are the properties of any black holes among the decay products? How do they differ from their progenitors, and what are the consequences for the classical restrictions on black hole parameters?

We will show that, in this case, the black holes produced by WGC decay do \emph{not} necessarily violate the classical Cosmic Censorship condition, if the decay involves a loss of angular momentum rather than charge, unless their masses are sufficiently small. They do however violate another, less well-known condition, namely the requirement that the specific angular momentum of the black hole should not exceed the asymptotic AdS length scale. As a consequence, they are strange objects indeed: the metric component corresponding to the colatitude changes sign when the colatitude is small, just as the metric component corresponding to the radial coordinate does when it is small. Their existence is nevertheless an (almost\footnote{The loophole is the possibility that ``WGC decay'' \emph{only} occurs through the emission of charged, non-spinning black holes. We regard such a prohibition on emission of rotating black holes as highly implausible, but cannot rule it out.}) inevitable consequence of the black hole version of the WGC; and this is the principal conclusion of the present work.

We begin with a brief survey of the properties of classical AdS$_4$-Kerr-Newman black holes.

\addtocounter{section}{1}
\section* {\large{\textsf{2. Basics of AdS$_4$-Kerr-Newman Black Holes}}}
We will discuss the case in which an asymptotically AdS$_4$ black hole rotates and is \emph{magnetically} charged \cite{kn:weinberg}, since this case currently attracts considerable attention, from both the observational \cite{kn:gustavo,kn:rong,kn:mota,kn:ronggen,kn:diamond,kn:tang} (see \cite{kn:yang,kn:ullah} for reviews) and theoretical \cite{kn:juan0,kn:juan1,kn:juan2,kn:103} points of view. The magnetic case is particularly interesting to us because such a black hole is expected to be, generically, approximately extremal \cite{kn:juan1}, and extremal black holes are of course the ones of most interest in connection with the WGC\footnote{However, in the present work (though not always in \cite{kn:103}), the magnetic black hole parameter $P$ can always be replaced by $\sqrt{P^2 + Q^2}$, where $Q$ is the electric black hole parameter, so there is no loss of generality in focussing on this case.}.

In the case where the $r = $ constant sections have the topology of a sphere, the four-dimensional magnetic AdS-Kerr-Newman metric \cite{kn:cognola} has the form
\begin{flalign}\label{A}
g(\m{AdS}_{a,M,P,L}) = &- {\Delta_r \over \rho^2}\Bigg[\,\m{d}t \; - \; {a \over \Xi}\sin^2\theta \,\m{d}\phi\Bigg]^2\;+\;{\rho^2 \over \Delta_r}\m{d}r^2\;+\;{\rho^2 \over \Delta_{\theta}}\m{d}\theta^2 \\ \notag \,\,\,\,&+\;{\sin^2\theta \,\Delta_{\theta} \over \rho^2}\Bigg[a\,\m{d}t \; - \;{r^2\,+\,a^2 \over \Xi}\,\m{d}\phi\Bigg]^2,
\end{flalign}
where
\begin{eqnarray}\label{B}
\rho^2& = & r^2\,+\,a^2\cos^2\theta, \nonumber\\
\Delta_r & = & (r^2+a^2)\Big(1 + {r^2\over L^2}\Big) - 2Mr + {P^2\over 4\pi},\nonumber\\
\Delta_{\theta}& = & 1 - {a^2\over L^2} \, \cos^2\theta, \nonumber\\
\Xi & = & 1 - {a^2\over L^2}.
\end{eqnarray}
Here $L$ is the asymptotic AdS$_4$ curvature scale, $a$ is the angular momentum per unit physical mass, and $M$ and $P$ are parameters related \cite{kn:gibperry} to the physical mass $\mathcal{M}$ and the physical magnetic charge $\mathcal{P}$ by
\begin{equation}\label{C}
\mathcal{M}\;=\;M/(\ell_{\textsf{G}}^2\Xi^2), \;\;\;\;\;\mathcal{P}\;=P/(\ell_{\textsf{G}}\Xi),
\end{equation}
where $\ell_{\textsf{G}}$ is the gravitational length scale\footnote{For reasons that are clear from these relations, we do not allow $a$ to be precisely equal to $L$. That might be possible in some limiting sense \cite{kn:superent}, but the resulting objects are quite different from those we study here.}. Notice that these equations mean that the physical mass and charge actually depend on the angular momentum, even if the geometric black hole parameters $M$ and $P$ are fixed. This is a serious complication in dealing with these black holes.

The (outer) horizon is located at $r = r_{\textsf{H}}$, obtained by solving the equation
\begin{equation}\label{D}
\Delta_r(r_{\textsf{H}})\;=\;(r_{\textsf{H}}^2+a^2)\Big(1 + {r_{\textsf{H}}^2\over L^2}\Big) - 2Mr_{\textsf{H}} + {P^2\over 4\pi}\;=\;0
\end{equation}
for its largest root. This equation is difficult to solve in the general case, but, by using the fact that the Hawking temperature vanishes for extremal black holes, one can give a simple expression for $r_{\textsf{H}}$ in the extremal case:
\begin{equation}\label{E}
r_{\textsf{H}}^2 \;=\; {L^2\over 6}\left(-1 - {a^2\over L^2} + \sqrt{\left(1 + {a^2\over L^2}\right)^2 + {12\over L^2}\left(a^2\,+\,{P^2\over 4\pi}\right)}\right).
\end{equation}
Notice that this expression does not involve $M$. Therefore, requiring the existence of real solutions of equation (\ref{D}) gives a lower bound for $M$ in terms of $a, P,$ and $L$. This is of course the Cosmic Censorship bound, but it is given in terms of the geometric parameters, not the physical mass and charge.

To remedy this, we use (\ref{C}), and then we have
\begin{equation}\label{F}
r_{\textsf{H}}^2 \;=\; {L^2\over 6}\left(-1 - {a^2\over L^2} + \sqrt{\left(1 + {a^2\over L^2}\right)^2 + {12\over L^2}\left(a^2\,+\,{\ell_{\textsf{G}}^2\left(1 - [a^2/L^2]\right)^2\mathcal{P}^2\over 4\pi}\right)}\;\right).
\end{equation}
Putting this back into (\ref{D}), and using (\ref{C}) again, we have the condition for Cosmic Censorship to be satisfied:
\begin{equation}\label{G}
\mathcal{M} \;\geq \; {\sqrt{6}L \left[{1\over 36} \left({5a^2\over L^2}\, -\, 1\, + \textsf{Z} \;\right) \times \left(5\, -\, {a^2\over L^2}\, +\, \textsf{Z} \;\right) + \,{\ell_{\textsf{G}}^2\left(1 - [a^2/L^2]\right)^2\mathcal{P}^2\over 4\pi L^2} \;\right]\over 2\ell_{\textsf{G}}^2\left(1 - [a^2/L^2]\right)^2\sqrt{-1\, -\, {a^2\over L^2} \,+\,\textsf{Z} }},
\end{equation}
where
\begin{equation}\label{GAG}
\textsf{Z}\;=\;\sqrt{\left(1 + {a^2\over L^2}\right)^2 + {12\over L^2}\left(a^2\,+\,{\ell_{\textsf{G}}^2\left(1 - [a^2/L^2]\right)^2\mathcal{P}^2\over 4\pi}\right)}.
\end{equation}

The graphs of the lower bound on $\mathcal{M}$ as a function of $a$ and $\mathcal{P}$ are shown\footnote{We are using units in which physical mass has units of inverse length, and charge and angular momentum are dimensionless. The axes on the figures correspond to the three dimensionless quantities $a/L, \; \ell_{\textsf{G}}\mathcal{P}/L, \; \mathcal{M}\ell_{\textsf{G}}^2/L.$} as Figure 1 (for $a/L < 1$) and Figure 2 (for $a/L > 1.$)
\begin{figure}[!h]
\centering
\includegraphics[width=0.9\textwidth]{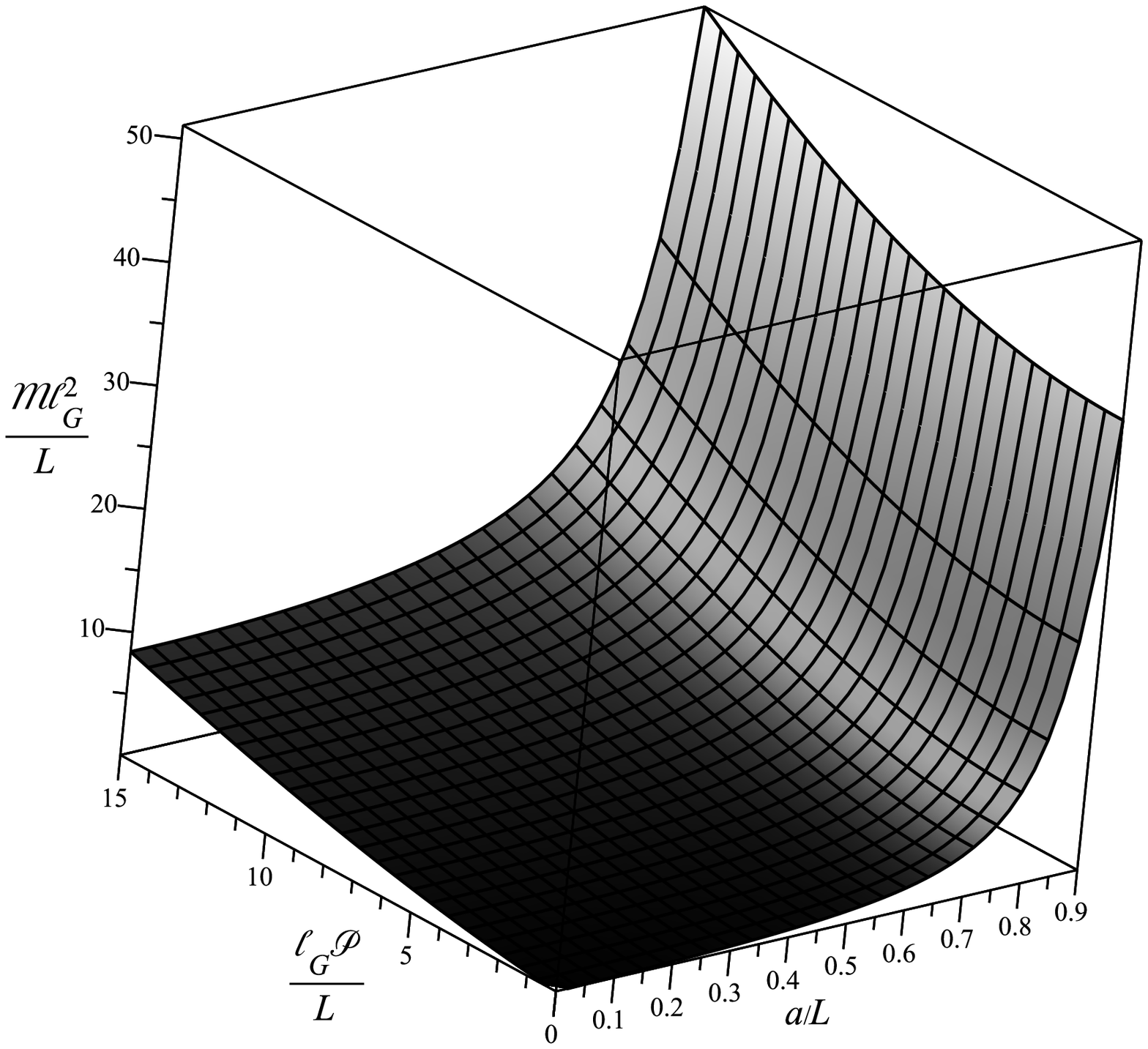}
\caption{AdS$_4$-Kerr-Newman black holes with $a/L < 1$ satisfying Cosmic Censorship correspond to points on or above the surface shown.}
\end{figure}
\begin{figure}[!h]
\centering
\includegraphics[width=0.9\textwidth]{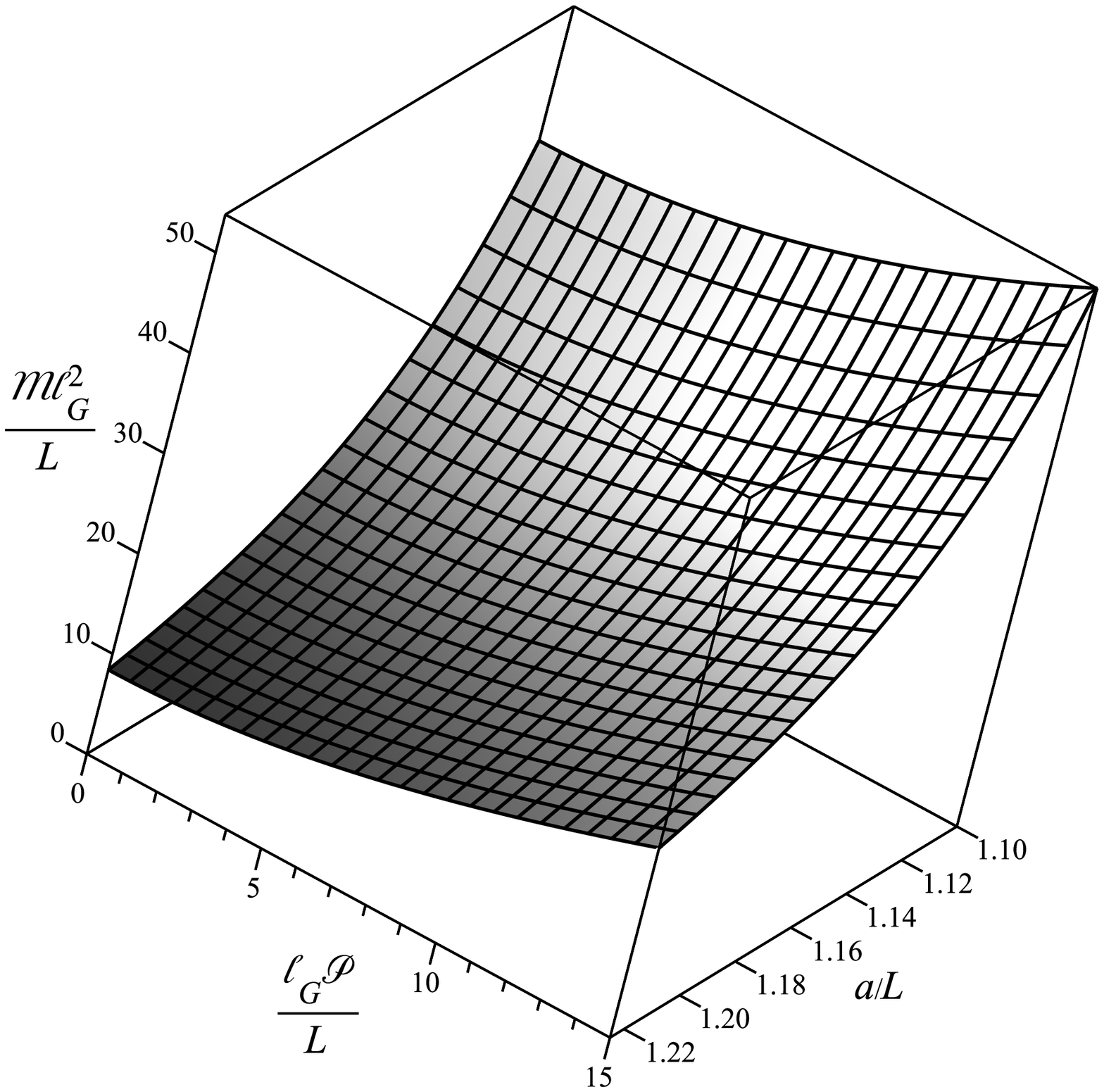}
\caption{AdS$_4$-Kerr-Newman black holes with $a/L > 1$ satisfying Cosmic Censorship correspond to points on or above the surface shown.}
\end{figure}
It is important to note that both graphs are convex surfaces. The physical meaning of this in the $a/L < 1$ case is that, for a given charge, the Censorship condition is more restrictive in the asymptotically AdS$_4$ case than for its asymptotically flat counterpart: that is, the mass has to be larger if Censorship is to hold for a black hole of given charge.

The Taylor expansion for the right side of (\ref{G}) as a function of $\mathcal{P}$ for $a = 0$ gives us
\begin{equation}\label{GG}
\mathcal{M} \;\geq \;{\mathcal{P}\over 2\sqrt{\pi}\ell_{\textsf{G}}} \,+\, {\ell_{\textsf{G}}\over 16\pi^{3/2}L^2}\mathcal{P}^3 \,+\, \mathcal{O}(\mathcal{P}^5),
\end{equation}
while that for the right side as a function of $a$ for $\mathcal{P} = 0$ gives
\begin{equation}\label{GGG}
\mathcal{M}\; \geq \;{a\over \ell_{\textsf{G}}^2} \,+\, {3\over \ell_{\textsf{G}}^2L^2}a^3 \,+\, \mathcal{O}(a^5).
\end{equation}
We see that the Censorship conditions in the asymptotically flat cases ($\mathcal{M} \geq \mathcal{P}/(2\sqrt{\pi}\ell_{\textsf{G}})$ for Reissner-Nordstr\"{o}m, $\mathcal{M} \geq a/\ell_{\textsf{G}}^2$ for Kerr) are recovered in the limit $L \rightarrow \infty.$ Again, the forms of the expansions reflect the fact that the Censorship condition is more restrictive in the asymptotically AdS$_4$ case.

For any given finite $\mathcal{M},$ it is clear (from the prefactor on the right) that the condition (\ref{G}) can be violated by taking $a$ sufficiently \emph{close} to $L$. However, this in itself does not mean that $a$ cannot \emph{exceed} $L$: it can, provided that the right side of (\ref{G}) is well-defined when $a/L$ is strictly larger than unity. From Figure 2, this is evidently so. Thus, classical Cosmic Censorship does not rule out $a/L > 1,$ provided that $a/L$ is sufficiently \emph{large} in that case\footnote{The fact that Censorship cannot itself enforce $a/L < 1$ is related to the fact that the latter condition does not involve $\ell_{\textsf{G}}$, whereas of course Censorship does.}. In other words, classical Censorship only excludes a \emph{band} of values for $a/L$, on both sides of unity.

Let us consider the case of $a/L > 1$ in more detail\footnote{The (very different) five-dimensional AdS-Kerr case where $a/L > 1$ was discussed in \cite{kn:96}.}.

\addtocounter{section}{1}
\section* {\large{\textsf{3. AdS$_4$-Kerr-Newman Black Holes with $a/L > 1$.}}}
As usual, there is a coordinate ``singularity'' at the horizon of such a black hole, where $\Delta_r = 0.$ In exactly the same way, when $a/L > 1,$ there is an apparent ``singularity'' where $\Delta_{\theta} = 0$. That occurs at a specific value of $\theta$, $\theta_{\textsf{H}},$ where
\begin{equation}\label{H}
\theta_{\textsf{H}}\;=\;\arccos(L/a).
\end{equation}

As might be expected, the locus $\theta = \theta_{\textsf{H}}$ is also a coordinate ``singularity''. The simplest way to see this is to project out to conformal infinity ($r \rightarrow \infty$). One can evaluate the square of the curvature tensor for this three-dimensional space, and the result is
\begin{equation}\label{I}
R^{ijkl}R_{ijkl}\;=\;{4\over L^4}\,\left[1 + {a^4\over L^4} + 9\,{a^4\over L^4}\cos^4\theta - 4\,{a^2\over L^2}\cos^2\theta\left(1 + {a^2\over L^2}\right)\right];
\end{equation}
clearly this diverges nowhere, and the same is true of all other curvature invariants: there is no true singularity at $\theta = \theta_{\textsf{H}}$.

A useful way of characterising this locus is to note that the (outer) ergosphere of such a black hole, described by solving the equation
\begin{equation}\label{II}
-\,\Delta_r\;+\;a^2\sin^2\theta\,\Delta_{\theta}\;=\;0
\end{equation}
for $r$ as a function of $\theta$, actually intersects the outer horizon there (as well as at the pole, as usual). One finds that the outer ergosphere is, as usual, outside the outer horizon for $\theta > \theta_{\textsf{H}}$, but plunges below it at $\theta = \theta_{\textsf{H}}$ and remains under it for all $\theta$ with $0 < \theta < \theta_{\textsf{H}}.$ (Conversely, the inner ergosphere intersects the inner horizon at $\theta = \theta_{\textsf{H}}$, but it lies outside the inner horizon in $0 < \theta < \theta_{\textsf{H}}.$)

Now consider a point with $\theta > \theta_{\textsf{H}}$, and outside the outer ergosphere, so that all of the coordinates have their usual characters. Now move towards smaller values of $\theta$, keeping $r$ fixed (so that, as explained above, we remain outside the outer ergosphere). When $\theta = \theta_{\textsf{H}}$ is reached, $\m{d}\theta$ becomes null, exactly as $\m{d}r$ does when the event horizon is crossed. The points with $\theta = \theta_{\textsf{H}}$ therefore constitute a kind of ``angular horizon'': $\theta$ changes character at small\footnote{We stress that the exotic behaviour occurs in a neighbourhood of the \emph{poles}, not the equator, so it is not directly related to the fact that the black hole is ``rotating very rapidly''.} values, just as $r$ does.

The physical interpretation of such a spacetime is not completely clear. One can expect closed timelike curves to be present, but this is a common feature of rotating spacetimes, like the G\"{o}del spacetime; whether it is acceptable continues to be debated (see for example \cite{kn:kip,kn:matt,kn:caldaklemm,kn:calda}; see also \cite{kn:rov} for a discussion in relation to the second law of thermodynamics). Notice that, inside the angular horizon, the space at conformal infinity in this case has (effectively) Euclidean signature $(-, -, -)$, like the conformal infinity of de Sitter spacetime, so perhaps the correct interpretation of the interior is in terms of a suitable adaptation of the ``dS/CFT correspondence'' \cite{kn:strom,kn:maldy}. We will not pursue this here: the point is that it is arguable that AdS$_4$-Kerr-Newman black holes with $a/L > 1$ should not be dismissed out of hand, though they are certainly exotic.

There are in fact more general hints in this direction. For example, the Euclidean versions of these spacetimes certainly allow $a/L > 1$, since, in passing to the Euclidean section we need to complexify $a$, with the result that the Euclidean versions of $\Delta_{\theta}$ and $\Xi$ are strictly positive for all values of the parameters. Thus, forbidding $a > L$ in the Lorentzian case would entail having an infinite family of Euclidean ``black hole'' geometries with no Lorentzian counterparts. Again, as we saw, values of $a/L$ greater than unity can be compatible with Cosmic Censorship: we just need to avoid values too close to unity.

More generally, the parameter $a$ is a local characteristic of the black hole, whereas $L$ describes the asymptotic geometry. From this perspective there is no obvious physical reason to expect one parameter to constrain the other. Such a constraint might arise from an embedding of General Relativity into a more complex theory containing extended objects, capable of probing large-scale geometry, which the black hole might emit. String theory is such a theory. We will see later that precisely this does indeed occur when these black holes are regarded as string-theoretic objects. This constraint, however, turns out not to be $a/L < 1$ but rather $a/L < \sqrt{3/2} \approx 1.225.$ The very fact that string theory \emph{does} allow a narrow range of values for $a$ above $L$ can be regarded as a hint that such values may be physical under some circumstances.

Of course, all this is in any case moot unless we can propose a concrete mechanism for actually producing such objects. It is far from clear that such a mechanism exists or can exist. For example, these black holes cannot be obtained (without violating Cosmic Censorship) simply by ``spinning up'' a black hole with $a/L < 1$: the two parameter ranges are disjoint, physically as well as mathematically. Again, one does not expect to obtain such objects by means of classical evolution from initial data.

Nevertheless, as we will see, a non-classical mechanism for producing black holes with $a/L > 1$ does in fact exist, so their possible existence will have to be confronted at some point.

In summary, it is not clear that $a < L$ is necessary, though certainly AdS$_4$-Kerr-Newman black holes with $a > L$ are very unusual\footnote{See also \cite{kn:101}.}. However, we will soon see that, if it nevertheless turns out to be the case that such objects must for some reason be rejected, the consequences will be very serious.

We now set about showing that, if the black hole version of the WGC is valid, then these exotic AdS$_4$-Kerr-Newman black holes must exist. The first point to understand here is that, in order for near-extremal rotating AdS$_4$ black holes to be stable, even classically, they must have a non-zero charge (in our case, a magnetic charge). Let us briefly review this point.

\addtocounter{section}{1}
\section* {\large{\textsf{4. Avoiding Superradiance }}}
It was shown in \cite{kn:hawkreall} that rotating asymptotically AdS$_4$ black holes can avoid a superradiant instability \cite{kn:super} if their angular velocities satisfy a condition which can be expressed in the form
\begin{equation}\label{J}
\xi \; \equiv \; {aL\,\left(1 + {r_{\textsf{H}}^2\over L^2}\right)\over r_{\textsf{H}}^2 + a^2}\;\leq\;1.
\end{equation}
This condition, which applies also in the charged case, amounts to a requirement that it should be possible to establish a thermal equilibrium of the black hole with co-rotating thermal radiation. (In terms of the AdS/CFT duality, it is the requirement that the matter described by the CFT at conformal infinity must not rotate ``faster than light''.) There is evidence that this condition is necessary as well as sufficient for stability in this sense \cite{kn:reall}.

Extremal, uncharged AdS$_4$-Kerr black holes with $a/L < 1$ always violate this condition\footnote{For the asymptotically flat case see \cite{kn:lin}.}: see Figure 3.
\begin{figure}[!h]
\centering
\includegraphics[width=0.9\textwidth]{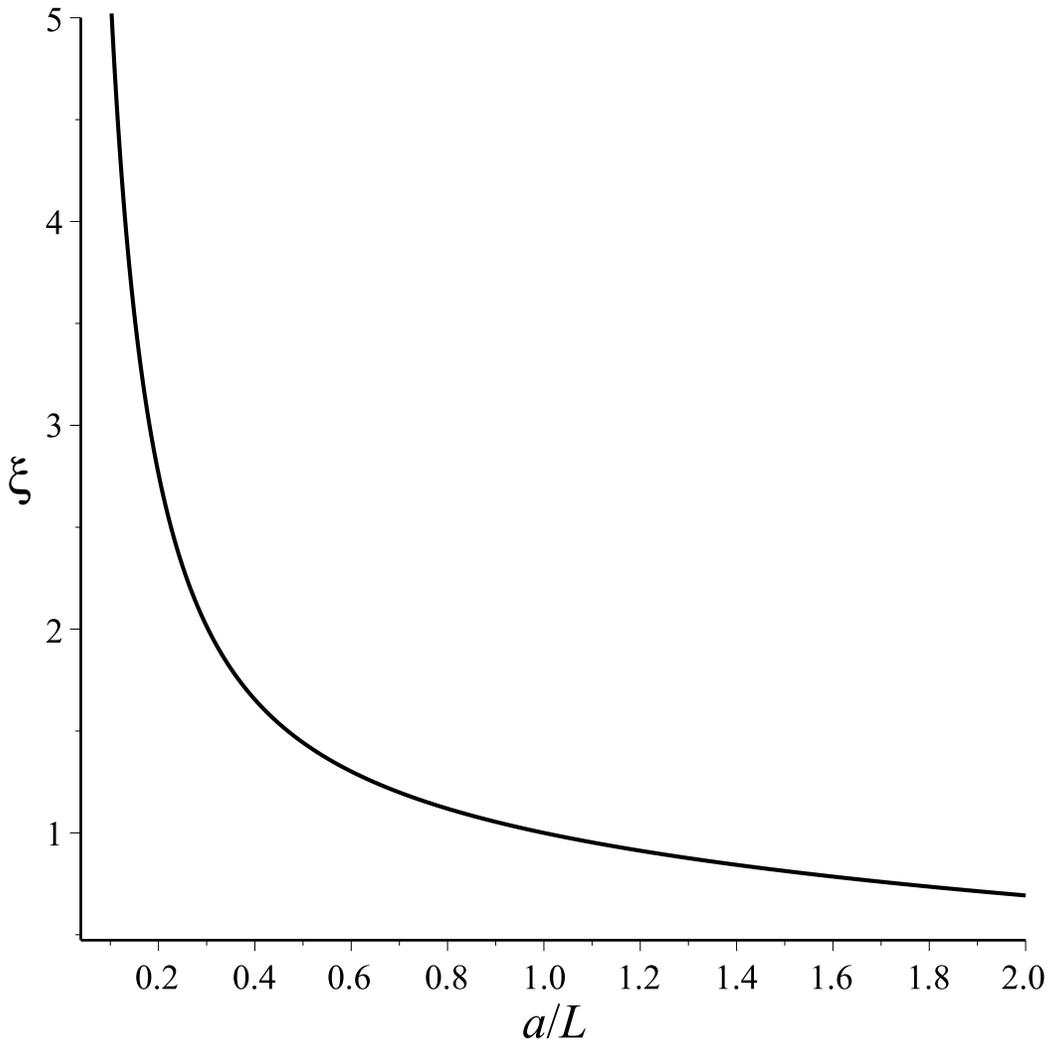}
\caption{The dimensionless function $\xi(a/L)$ for neutral, extremal AdS$_4$-Kerr black holes. It is greater than unity for all $a/L < 1,$ smaller than unity for all $a/L >1.$}
\end{figure}
Thus, they are always unstable \cite{kn:reall}. It is therefore extremely difficult to assess how such black holes might behave in the context of the WGC. Fortunately, one can reduce $\xi$ to below unity simply by adding electric or magnetic charge. For a given value of the angular momentum parameter $a,$ it was found in \cite{kn:103} that the minimal value of the charge needed to ensure stability against superradiance for an extremal black hole is given by a surprisingly simple expression:
\begin{equation}\label{K}
\mathcal{P} \;\geq\;\mathcal{P}_{\textsf{min}}\;=\;{2\,\sqrt{\pi a L}\over \ell_{\textsf{G}}|\left(1-{a\over L}\right)|}.
\end{equation}
In short, rotating extremal black holes in AdS$_4$ with $a/L < 1$ can be classically stable if their charge is sufficiently large. (By contrast, neutral extremal black holes with $a/L > 1$ are \emph{always} stable against superradiance.)

Notice that $\mathcal{P}_{\textsf{min}}$ is an increasing function of $a$ when $a/L < 1.$ It follows that, if the charge on an extremal AdS$_4$-Kerr-Newman black hole with $a \neq 0$ but $a/L < 1$ exceeds this value only slightly, then the emission of a massive, charged, but non-rotating object will increase $a$ (the angular momentum per unit mass) and therefore $\mathcal{P}_{\textsf{min}}$; but it will decrease\footnote{It cannot increase the charge if the resulting object is to satisfy classical Cosmic Censorship.} $\mathcal{P}$, resulting in some cases in a violation of (\ref{K}). This will result in an instability if the surviving black hole is extremal or sufficiently close to extremality. In short, ``WGC instability'' can trigger a superradiant instability in some cases.

\addtocounter{section}{1}
\section* {\large{\textsf{5. String-Theoretic Considerations }}}
As was mentioned earlier, it seems strange that the angular momentum parameter $a$, which (with $\mathcal{M}$ and $\mathcal{P}$) controls the geometry near to the black hole, should be related in any way to the parameter $L$ that controls the asymptotic geometry. This was explained in \cite{kn:96} for five-dimensional rotating AdS black holes in string theory; we refer the reader there for a full discussion. Here we will focus on the ways in which the four-dimensional case differs from the situation in five dimensions.

The point is that string theory allows (and requires) us to probe the black hole geometry with extended objects, branes. As these branes propagate away from the black hole towards infinity, they carry information about the black hole parameters, including $a$. It turns out that, at large distances from the black hole, $a$ and $L$ have competing influences on the brane physics, and their ratio determines whether the brane action in that region is physical (that is, bounded below). If it is not, then the system does not exist as a well-defined solution in string theory. This was first discussed in the general case by Seiberg and Witten \cite{kn:seiberg}; see also \cite{kn:maldacena,kn:KPR}.

The action of a BPS brane located at a fixed value of $r$ in the AdS$_4$-Kerr-Newman geometry takes the form (calculated in much the same manner\footnote{The five-dimensional case differs in two ways: first, there are straightforward differences in the coefficients because of the different number of dimensions, and second, in the five-dimensional case $a$ does not have a direct physical interpretation as the ratio of the black hole angular momentum to its mass.} as in the five-dimensional case discussed in \cite{kn:96})
\begin{equation}\label{L}
\mathfrak{S}\;\propto\;r\,\sqrt{\Delta_r}\left[\sqrt{1 + {a^2\over r^2}} + {r\over a}\,\mathrm{arcsinh}{a\over r}\right]\;-\;{2r^3\over L}\left(1 + {a^2\over r^2}\right) \;+\;{2r_{\textsf{H}}^3\over L}\left(1 + {a^2\over r_{\textsf{H}}^2}\right),
\end{equation}
with a positive overall constant of proportionality.

At large values of $r$, one has
\begin{equation}\label{M}
\mathfrak{S}\;\propto\;\left(1 - {2a^2\over 3L^2}\right)rL \;+\;\mathcal{O}(1/r).
\end{equation}
We see at once that the action will be bounded below at large $r$ provided that $a$ and $L$ satisfy
\begin{equation}\label{N}
a/L \;\leq\;\sqrt{3/2}\, \approx \, 1.225.
\end{equation}

We conclude that AdS$_4$-Kerr-Newman black holes exist as solutions of string theory only when (\ref{N}) is satisfied. When the discussion is embedded in string theory, then, it becomes possible to provide a definite physical justification for putting a bound on $a/L$. We note, however, that the upper bound is not unity ---$\;$ though it is not far above it. One would expect that, if a mechanism for producing black holes with $a/L > 1$ can be found, then they will be produced with a wide range of values for $a/L$, and we will see later that this is correct. All of those with $a/L > \sqrt{3/2}$ will have to shed angular momentum by means of brane pair-production \cite{kn:maldacena}, until (\ref{N}) is restored. We therefore expect that the generic black hole of this sort, produced in this manner, will have $a/L \approx \sqrt{3/2}.$ We will be more precise about this later: it will turn out that this is indeed so, except possibly when the black hole is formed by being emitted by a larger black hole with a specific angular momentum which is relatively large (below but close to $L$).

We will see that the fact that string theory permits $a/L > 1$ plays an essential role in the physics of the WGC for AdS$_4$-Kerr-Newman black holes.

\addtocounter{section}{1}
\section* {\large{\textsf{6. Decay of Extremal AdS$_4$-Kerr-Newman Black Holes }}}
Let us begin with an AdS$_4$-Kerr-Newman black hole, with mass $\mathcal{M},$ magnetic charge $\mathcal{P}$, and angular momentum $\mathcal{J}$. (For this black hole, we will denote the angular momentum parameter by $\mathcal{A},$ so $\mathcal{A} = \mathcal{J}/\mathcal{M}.$ When using the Censorship inequality (\ref{G}), we simply replace $a$ with $\mathcal{A}.$) Suppose that this black hole is sufficiently massive that, to a good approximation, it satisfies (and continues to satisfy, under small perturbations) classical Cosmic Censorship.

We now assume that this black hole emits a much smaller AdS$_4$-Kerr-Newman black hole with physical parameters $m, p, j,$ which for simplicity we take to be all positive. (We will use $a$ for the angular momentum parameter of this black hole, that is, $a = j/m.$) The parameters of the original black hole become $\mathcal{M} + \delta\mathcal{M}, \mathcal{P} + \delta\mathcal{P}, \mathcal{J} + \delta\mathcal{J}.$

Allowing for the possibility that the smaller black hole is moving (radially) relative to the original black hole, we have
\begin{equation}\label{P}
m \;\leq\; -\,\delta\mathcal{M},
\end{equation}
while the conservation of charge requires
\begin{equation}\label{Q}
p = -\,\delta \mathcal{P}.
\end{equation}
To avoid complications arising from the anisotropy of the spacetime (in particular, from frame-dragging), we assume for simplicity that the smaller black hole is emitted along the axis of symmetry, so that it has no orbital angular momentum. The conservation of angular momentum is then expressed by
\begin{equation}\label{R}
j = -\,\delta \mathcal{J}.
\end{equation}
Thus we have
\begin{equation}\label{S}
{m\over p} \;\leq\; {\delta\mathcal{M}\over \delta\mathcal{P}}.
\end{equation}
and likewise
\begin{equation}\label{T}
{m\over j} \;\leq\; {\delta\mathcal{M}\over \delta\mathcal{J}}.
\end{equation}

Suppose now that the original, ``large'' black hole satisfies $\mathcal{A}/L < 1$, is extremal, and continues to obey classical Cosmic Censorship as it emits the smaller black hole. (There are of course strong reasons to believe this in the case of decay by Hawking evaporation, both with regard to electromagnetic charges (see \cite{kn:ong} for a recent comprehensive discussion of Hawking radiation and its relation to the WGC) and angular momentum \cite{kn:page}.)

This means that, as the parameters change, $\mathcal{M}$ must not decrease with any more rapidly than the inequality (\ref{G}) permits. For simplicity, we will consider the cases in which only one parameter changes at a time; that is, the emitted black hole has either zero angular momentum or zero charge\footnote{It is possible that stronger conclusions could be reached by allowing both parameters to change simultaneously. In view of the structure of the conclusion we reach below (inequality (\ref{DELTA})), this seems very unlikely, but the question needs further consideration.}. Then we have, respectively,
\begin{equation}\label{U}
{\delta\mathcal{M}\over \delta\mathcal{P}} \;\leq\; {\partial\mathcal{M}\over \partial\mathcal{P}}
\end{equation}
and
\begin{equation}\label{V}
{\delta\mathcal{M}\over \delta\mathcal{J}} \;\leq\; {\partial\mathcal{M}\over \partial\mathcal{J}}.
\end{equation}
Combining (\ref{S}) with (\ref{U}) and (\ref{T}) with (\ref{V}) we have
\begin{equation}\label{W}
{m\over p} \;\leq\; {\partial\mathcal{M}\over \partial\mathcal{P}}
\end{equation}
and
\begin{equation}\label{X}
{m\over j} \;\leq\; {\partial\mathcal{M}\over \partial\mathcal{J}}.
\end{equation}
Note that, in the asymptotically AdS$_4$ case (unlike for its asymptotically flat counterpart), the derivatives here are \emph{not constants}. Thus, if we want it to be possible for \emph{every} classically stable extremal black hole to decay, we have to allow for this and ensure that these inequalities are satisfied at some point or points where the derivatives are minimal. Furthermore, we must interpret ``minimal'' to mean, ``minimal on some \emph{physically acceptable} domain''. All this greatly complicates our task.

Let us begin with (\ref{W}). As mentioned earlier, $\mathcal{M}$ is a convex function of $\mathcal{P}$ and $a$. We can therefore minimise the derivative with respect to $\mathcal{P}$ by setting $a = 0$; this is obviously compatible with (\ref{K}).  The minimal possible value of the derivative occurs at $\mathcal{P} = 0,$ and can be read off from equation (\ref{GG}), so we conclude that every extremal AdS$_4$-Kerr-Newman black hole can decay (to a near-extremal black hole which satisfies Cosmic Censorship) by emitting a zero-spin object if and only if that object satisfies (inverting the inequality to put it in a more familiar form)
\begin{equation}\label{Y}
{p \over m} \;\geq\; 2\sqrt{\pi}\ell_{\textsf{G}}.
\end{equation}
This is in fact precisely the same\footnote{We use Lorentz-Heaviside units.} as the familiar statement of the WGC in the asymptotically flat case \cite{kn:motl,kn:palti,kn:rude,kn:rem,kn:shiu,kn:goon}. This has the well-known consequence: either the emitted object, if it is a small black hole (for which quantum corrections are important), violates Cosmic Censorship or, more interestingly, the quantum-corrected version of Cosmic Censorship must be such that values of $p/m$ satisfying (\ref{Y}) are compatible with it \cite{kn:kats}.

Now let us turn to the case (inequality (\ref{X})) where the emitted object carries angular momentum but no charge; the charge of the parent black hole will be fixed at some (necessarily non-zero) value compatible with (\ref{K}). (This is actually physically reasonable in the case of magnetically charged objects, which tend to be relatively very massive and therefore difficult to create \cite{kn:juan1}.)

Since $\mathcal{M}$ is given in (\ref{G}) as a function of $\mathcal{A}$, not $\mathcal{J},$ we need to express one derivative in terms of the other:
\begin{equation}\label{Z}
{\partial \mathcal{M}\over \partial \mathcal{J}} = {1\over \mathcal{A}\;+\;\left(\mathcal{M}/{\partial \mathcal{M}\over \partial \mathcal{A}}\right)}.
\end{equation}
It is convenient to use this to re-state (\ref{X}) in the form (again inverting the inequality)
\begin{equation}\label{ALPHA}
{a\over L} \;=\;{j\over mL} \;\geq\; {\mathcal{A}\over L}\;+\;{\mathcal{M}\over L\,{\partial \mathcal{M}\over \partial \mathcal{A}}} \;\equiv \;\textsf{K}(\mathcal{P},\;\mathcal{A}/L).
\end{equation}
The dimensionless function $\textsf{K}(\mathcal{P},\;\mathcal{A}/L)$ on the right can be evaluated explicitly (after differentiating $\mathcal{M}$ in (\ref{G})) with a computer algebra system, but, as one might have expected, it is in general extremely complex; it would be pointless to reproduce it here. Its graph is shown in Figure 4.
\begin{figure}[!h]
\centering
\includegraphics[width=0.9\textwidth]{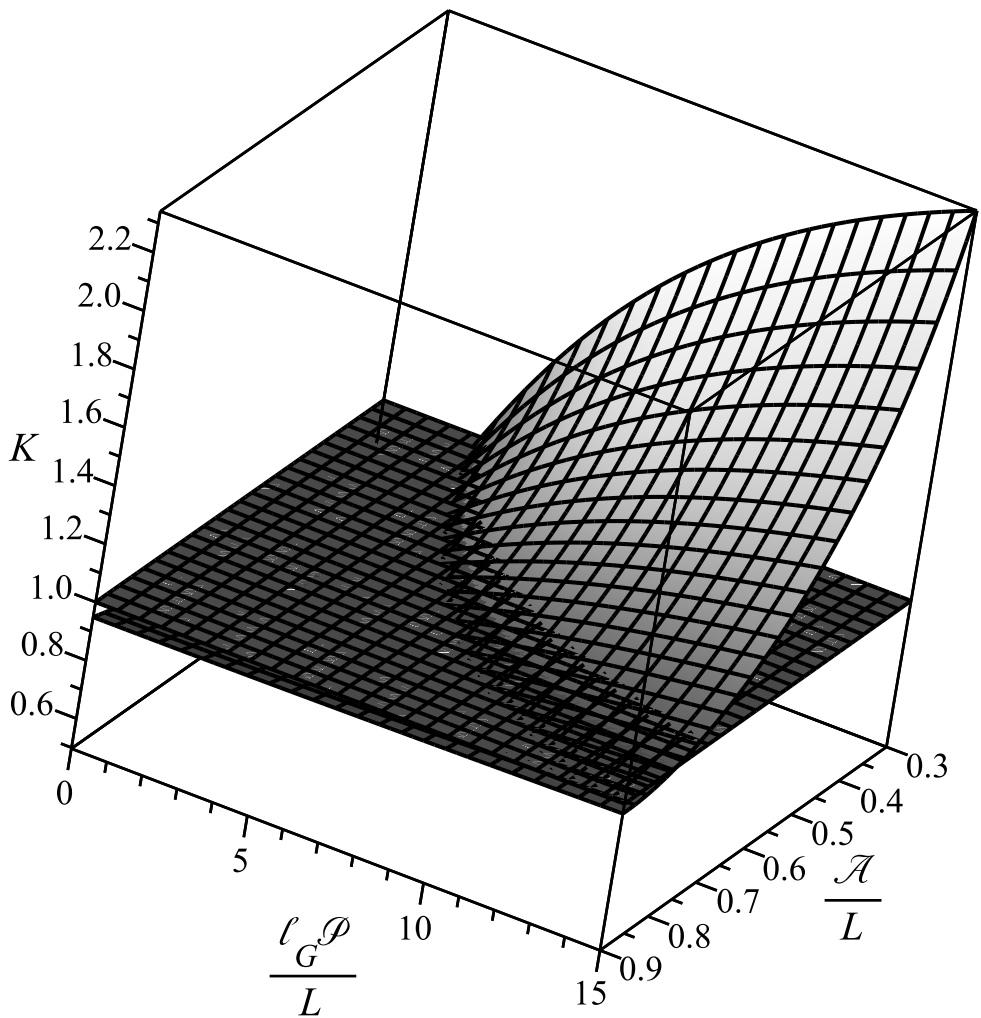}
\caption{The dimensionless function $\textsf{K}(\mathcal{P},\;\mathcal{A}/L)$ which gives a lower bound for $a/L$ for the emitted black hole. Also shown is $\textsf{K} = 1.$ The intersection curve is precisely the same as the curve in Figure 5 below. We see that requiring $\mathcal{P} \geq \mathcal{P}_{\textsf{min}}$ implies that $\textsf{K}(\mathcal{P},\;\mathcal{A}/L) \geq 1.$}
\end{figure}

Since $\mathcal{M}$ is an increasing function of $\mathcal{A}$ (for black holes with $\mathcal{A}/L < 1,$ as we are assuming here ---$\,$ see Figure 1), the derivative here is positive, and so we see that this relation implies that $a > \mathcal{A}.$

\emph{If} we assume that the emitted black hole satisfies $a/L < 1$, then, since the mass $m$ of the emitted black hole is smaller than that of the original black hole, and since the function in (\ref{G}) increases with $a$ when $a/L < 1,$ we see that $a > \mathcal{A}$ means that the emitted black hole necessarily violates classical Censorship. Notice that the assumption that $a/L < 1$ is crucial to this argument, because (see Figure 2) it is not the case that the function in (\ref{G}) increases with $a$ when $a/L > 1;$ on the contrary, it is strictly decreasing for such values of $a$.

This result is as expected, in the sense that the conclusion is the same as in the case where the emitted object is charged. As in the usual WGC arguments, we deduce that, \emph{if the emitted object satisfies} $a/L < 1,$ then Cosmic Censorship can hold for the emitted black hole only if the Censorship criterion is modified due to quantum effects \cite{kn:kats}.

But now we need to ask: does (\ref{ALPHA}) in fact require or even \emph{allow} $a/L < 1?$ The answer is that it does allow this, provided that $\mathcal{P}$ can be chosen freely: for, in that case, $\textsf{K}(\mathcal{P},\;\mathcal{A}/L)$ can be smaller than unity, as can be seen in Figure 4 (where the plane $\textsf{K}(\mathcal{P},\;\mathcal{A}/L) = 1$ has also been included: the graphs intersect). Then (\ref{ALPHA}) permits values of $a/L$ below unity.

But it is not the case that $\mathcal{P}$ can be chosen freely: as we saw above, the initial extremal black hole will not be stable, even classically, unless $\mathcal{P}$ satisfies the inequality (\ref{K}). Imposing that condition leads us to a remarkable conclusion.

First, note that, for fixed $\mathcal{A}/L < 1,$ the function $\textsf{K}(\mathcal{P},\;\mathcal{A}/L)$ is an increasing function of $\mathcal{P};$ see Figure 4. We therefore have
\begin{equation}\label{BETA}
\textsf{K}(\mathcal{P},\;\mathcal{A}/L)\; \geq \; \textsf{K}(\mathcal{P}_{\textsf{min}},\;\mathcal{A}/L).
\end{equation}
Now we propose to substitute (\ref{K}) into this inequality, and then use the result in (\ref{ALPHA}). It must be stressed that the form of the function $\textsf{K}(\mathcal{P},\;\mathcal{A}/L)$ \emph{does not depend in any way on any considerations related to superradiance}: it arises exclusively from the requirements imposed by the WGC. We therefore expect that substituting (\ref{K}) into the already extremely intricate function $\textsf{K}(\mathcal{P}_{\textsf{min}},\;\mathcal{A}/L)$ should yield an expression which is still more complicated and indeed unsurveyable.

And yet this is not the case: with the aid of a computer algebra system\footnote{The computations in this paper were performed by using Maple$^{\textsf{TM}}$.} we find that the expression simplifies dramatically, and the end result is simply
\begin{equation}\label{GAMMA}
\textsf{K}(\mathcal{P}_{\textsf{min}},\;\mathcal{A}/L)\;=\;1,
\end{equation}
for all values of $\mathcal{A}.$

This remarkable simplification seems to point to \emph{some deep relation between the WGC and superradiance.}

Leaving this observation to one side for the present, we see that, if the original, large black hole is stable against a superradiant instability, then the statement $a > \mathcal{A}$ can be greatly strengthened:
\begin{equation}\label{DELTA}
a/L\;=\;{j\over mL}\;>\;1,
\end{equation}
the inequality being strict because we do not allow $a/L = 1$ (see the relations (\ref{C})).

This can be seen graphically by examining the intersection of the graph of $\textsf{K}(\mathcal{P},\;\mathcal{A}/L)$ with that of $\textsf{K} = 1;$ this intersection can be seen in Figure 4. The resulting curve coincides precisely with the curve obtained by treating $\ell_{\textsf{G}}\mathcal{P}_{\textsf{min}}/L$ as a function of $a/L$ (as in the inequality (\ref{K})), as shown in Figure 5.

\begin{figure}[!h]
\centering
\includegraphics[width=0.8\textwidth]{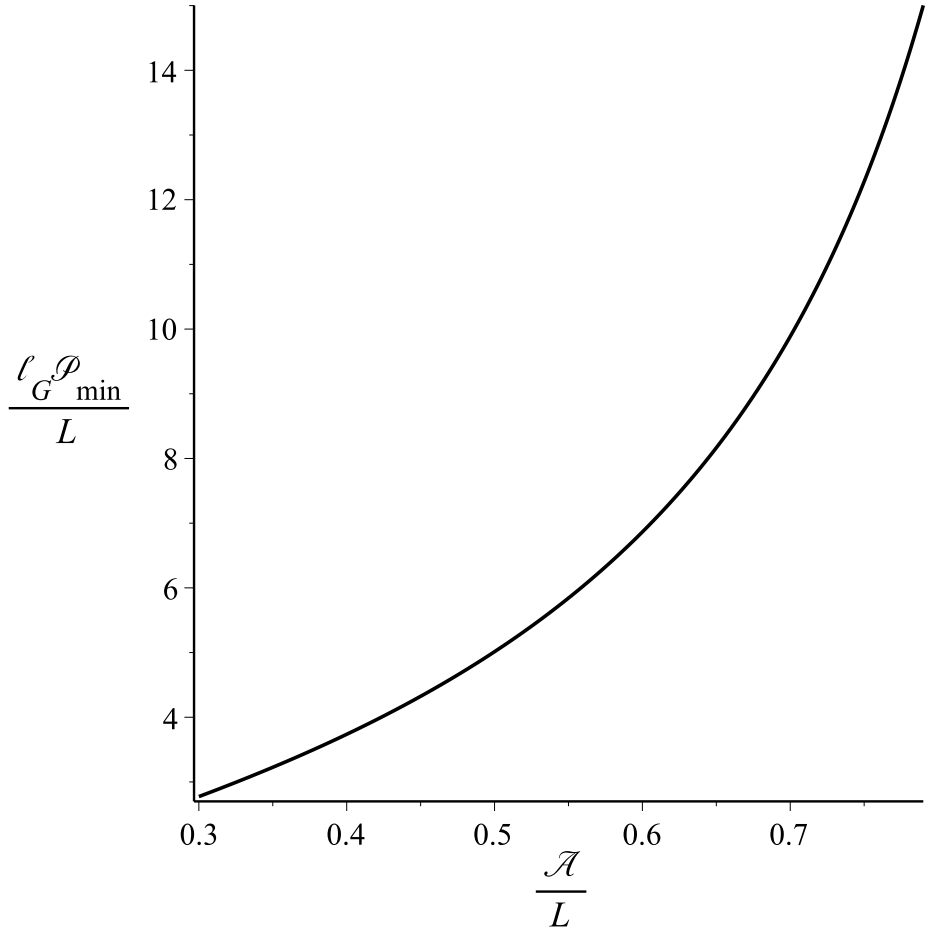}
\caption{The (scaled) minimal charge as a function of $\mathcal{A}/L$. The curve coincides with the intersection curve in Figure 4.}
\end{figure}

The conclusion is that the WGC demands that the emitted black hole \emph{must} be of the exotic variety we discussed earlier, with an ``angular horizon''. To put it another way: if such exotic black holes are deemed to be unphysical and their production therefore impossible, then the initial extremal black hole \emph{cannot} decay by emitting small rotating black holes; only decay by emission of charged, non-rotating black holes would be possible. Notice that the requirement that the original black hole should be stable against radiation of superradiant modes plays a crucial role here, because it implies that the initial black hole \emph{must} be charged. A neutral extremal AdS$_4$ black hole would be forced to decay by emitting rotating objects, but such an object cannot be stable.

Nevertheless, it seems unreasonable to posit such a strong asymmetry between charge and angular momentum. The original black hole is, by assumption, extremal, so it is rotating as rapidly as it can, given fixed values of the other parameters, without violating Cosmic Censorship. It would be strange indeed if, granted that it must decay, the black hole cannot do so by shedding some angular momentum.

Our point of view here is that WGC decay by emission of rotating black holes \emph{is} possible, and that in fact it offers an answer to the question: even if exotic black holes are physically possible, how can they be produced? The answer is that they are inevitably produced as extremal AdS$_4$-Kerr-Newman black holes decay, if that decay is via the emission of rotating black holes.

As we have seen, string theory does allow $a/L > 1$, but only up to a value just above unity. We argued earlier that the theory suggests that, generically, the inequality (\ref{N}) will be approximately saturated, that is, an exotic AdS$_4$-Kerr-Newman black hole produced in this manner will typically have $a/L \approx 1.225,$ unless the initial black hole has a large value of $\mathcal{A}/L$. We can now be more precise about this.

Figure 6 shows two curves. The lower curve is the same as in Figure 5: that is, in order for the initial black hole to be stable against emission of superradiant modes, its charge must correspond to a point above that curve. The upper curve is $\textsf{K}(\mathcal{P},\;\mathcal{A}/L) = \sqrt{3/2}.$ Any black hole with parameters corresponding to a point above this curve can, according to (\ref{ALPHA}), only decay by emitting black holes which are unstable to Seiberg-Witten emission of branes.
\begin{figure}[!h]
\centering
\includegraphics[width=0.8\textwidth]{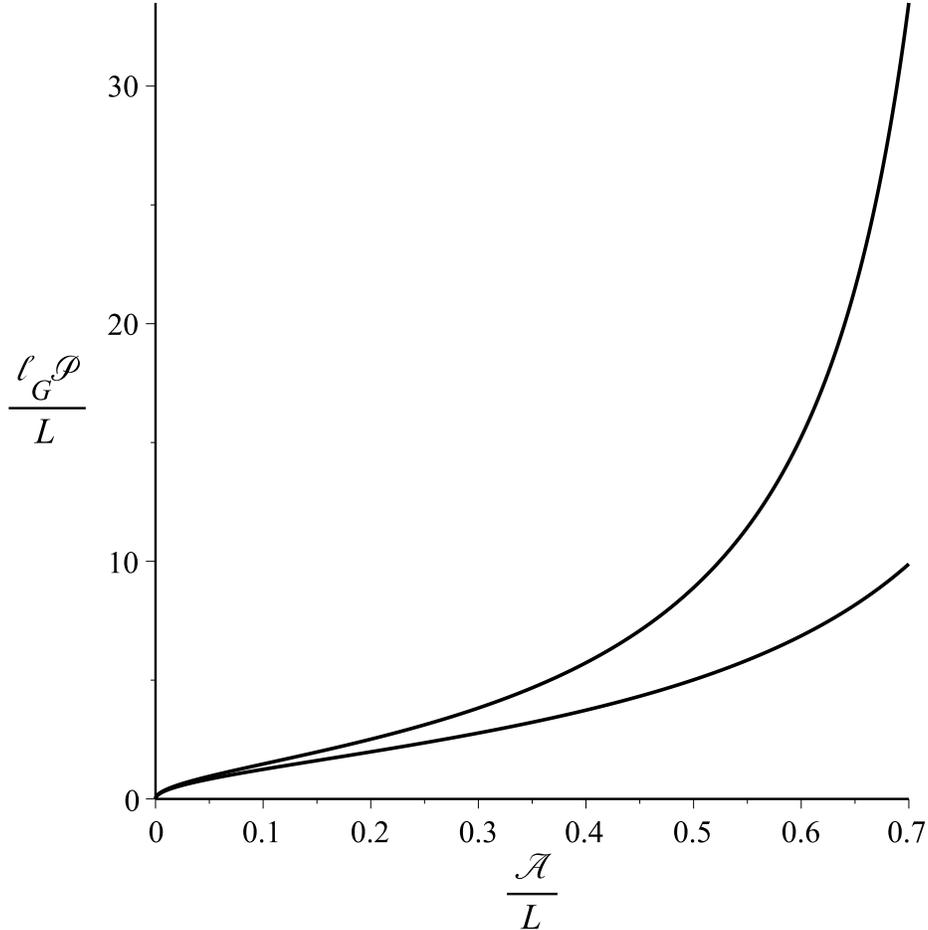}
\caption{Given a value for $\mathcal{A}/L,$ the emitted black hole will ultimately necessarily have $a/L \approx \sqrt{3/2}$ if $\ell_{\textsf{G}}\mathcal{P}/L$ takes a value above the upper curve; the emitting black hole will be unstable if $\ell_{\textsf{G}}\mathcal{P}/L$ takes a value below the lower curve. To the extent to which the gap between the curves is narrow, $a/L \approx \sqrt{3/2}$ is generic.}
\end{figure}
Such black holes will decay \cite{kn:maldacena} until $a/L \approx \sqrt{3/2}$ is approached. Therefore, once the emitted black hole settles down, it can only have $a/L \neq \sqrt{3/2}$ if the initial black hole has parameters corresponding to points strictly between these curves. For large values of $\mathcal{A}/L$, that is not difficult (indeed, it turns out that $\textsf{K}(\mathcal{P},\;\mathcal{A}/L)$ cannot attain $\sqrt{3/2}$ for any value of the charge if $\mathcal{A}/L \geq \sqrt{2/3} \approx 0.82$), but, for moderate values, it implies that only a narrow range of charges is available. For example, if $\mathcal{A}/L = 0.5,$ then $\ell_{\textsf{G}}\mathcal{P}/L$ has to lie between $\approx 5.013$ and $\approx 8.893$ for this to happen. This is the sense in which $a/L = \sqrt{3/2}$ is generic.

We can summarize the whole picture by stating that there is a kind of \emph{reciprocity} between the specific angular momenta of the two black holes: in a sense, the specific angular momentum of the emitted black hole is large ``because'' that of the original black hole is small. This is literally true in one case: the case where the charge is very large. For fixed $\mathcal{A},$ $\textsf{K}(\mathcal{P},\;\mathcal{A}/L)$ is a monotonically increasing function of $\mathcal{P},$ but it is bounded above by its limiting value; and (again) surprisingly, that limiting value has an extremely simple form:
\begin{equation}\label{DELTADELTA}
\lim_{\mathcal{P} \rightarrow \infty}\textsf{K}(\mathcal{P},\;\mathcal{A}/L)\; = \; L/\mathcal{A}.
\end{equation}
It follows that, to an approximation which can be made arbitrarily good by taking $\mathcal{P}$ to be sufficiently large, we have (from (\ref{ALPHA}))
\begin{equation}\label{DELTADELTADELTA}
{a\over L}\; \geq \; {L\over \mathcal{A}}.
\end{equation}
Of course, since we are assuming $\mathcal{A}/L < 1,$ (\ref{DELTA}) follows immediately from (\ref{DELTADELTADELTA}) when the charge is very large.

Notice that it follows from (\ref{DELTADELTADELTA}) and our earlier discussions that all highly charged AdS$_4$-Kerr-Newman black holes with $\mathcal{A}/L \leq \sqrt{2/3} \approx 0.82$ will emit small black holes which will eventually settle down to a state with $a/L \approx \sqrt{3/2}.$

Now let us turn to the question of the implications of our results for (classical) Cosmic Censorship.

\addtocounter{section}{1}
\section* {\large{\textsf{7. String Theory and Emitted Black Hole Masses }}}
At first sight, (\ref{DELTA}) seems to resemble the familiar dictates of the WGC: some black hole parameter (in this case, the angular momentum) is required to be larger, relative to the mass, than one would normally accept, as in the inequality (\ref{Y}) above. But here the situation is radically different: as we know, $a/L > 1$ does \emph{not} necessarily entail a violation of Cosmic Censorship. Its interpretation is quite different, involving the existence of an ``angular horizon'', leading us in an unexpected direction.

Before we abandon the idea that WGC decay of extremal AdS$_4$-Kerr-Newman black holes necessarily leads to violations of classical Cosmic Censorship, however, we should recall  \cite{kn:kats} that the WGC only requires \emph{substantial} deviations from classical Censorship (in the charged case) when the emitted object has such a low mass that quantum-gravitational effects cannot be completely neglected. We therefore need to ask whether our exotic black holes can continue to respect classical Censorship \emph{when their masses are small}. Part of the interest of such an investigation is that it might throw some light on the meaning of ``small'' masses in this context.

The precise nature of the emitted black hole in the above discussion is difficult to specify precisely; for example, it need not be extremal, and it may or may not satisfy classical Censorship. In this section, we ask the question: can the emitted black hole have an arbitrarily small mass and yet continue to satisfy classical Censorship? If not, what is the smallest possible mass for an emitted black hole which is stable and respects classical Censorship?

One sees from Figure 2 that black holes with $a/L > 1$ (which, in view of our discussion above, means all neutral rotating black holes emitted in accordance with the WGC from an extremal AdS$_4$-Kerr-Newman black hole) can respect classical Censorship even if their masses are very low, provided that $a/L$ can be very large. (The reader should recall that, in six or more dimensions, a black hole can have an arbitrarily large specific angular momentum without violating Cosmic Censorship \cite{kn:reall}, even in the asymptotically flat case. The fact that this is possible in the present case is therefore not unprecedented.)

The upshot is apparently that the emitted black holes can respect classical Cosmic censorship even with arbitrarily low masses. This argument depends, however, on the possibility of choosing arbitrarily large values for $a/L$. As we have seen, in string theory that is not an option if the emitted object is to be stable. The possible range of values of $a/L$ above unity is drastically curtailed: we have now $a/L < \sqrt{3/2}.$ Let us investigate the consequences.

The surface shown in Figure 2 describes the smallest possible dimensionless mass, as a function of $a/L$ and $\ell_{\textsf{G}}p/L$, of an exotic black hole which respects classical Censorship. Figure 2 has been drawn in such a way that values of $a/L$ greater than $\sqrt{3/2}$ are not shown. It is clear that, if the mass is too small, then exotic black holes cannot satisfy classical Cosmic Censorship and be stable in string theory.

The lowest point on this surface corresponding to a neutral black hole which satisfies classical Cosmic Censorship is described by a point with $a/L = \sqrt{3/2}$ and with $p = 0$. Using the dimensionless quantity $m\ell_{\textsf{G}}$ to measure the mass of the black hole as a multiple of the (AdS$_4$) Planck mass $1/\ell_{\textsf{G}}$, we find that all such black holes\footnote{We stress again that a black hole with mass saturating this inequality will be stable to superradiance, because it is extremal and uncharged, so, since $a/L > 1,$ it satisfies (\ref{K}) (Figure 3).} satisfy
\begin{equation}\label{ZETA}
m\ell_{\textsf{G}}\;\geq \; \approx 8.41 \, L/\ell_{\textsf{G}}.
\end{equation}

\emph{If the emitted black hole violates this bound, then it must also violate classical Cosmic Censorship}.

In short, when string-theoretic considerations are taken into account, sufficiently ``small'' neutral black holes emitted by an extremal AdS$_4$-Kerr-Newman black hole do violate classical Cosmic Censorship, where ``small'' is defined by (\ref{ZETA}).

In the AdS/CFT context, we can be somewhat more explicit, because the holographic ``dictionary'' expresses $L/\ell_{\textsf{G}}$ in terms of the rank of the dual gauge group, or, in the specific case of an asymptotically AdS$_4$ bulk \cite{kn:ABJM,kn:AdS4}, in terms of the number $N_{\textsf{c}}$ of coincident M2-branes in the dual field theory description: we have
\begin{equation}\label{ETA}
{L\over\ell_{\textsf{G}}} = {\left(2N_{\textsf{c}}\right)^{3/4}\over \sqrt{3}}.
\end{equation}
In a specific fully string-theoretic treatment, then, this ratio is (in principle) fixed, and the minimal mass in (\ref{ZETA}) can be computed.

In applications of the AdS/CFT duality, the ratio $L/\ell_{\textsf{G}}$ is normally assumed to be ``large'', so that a classical background can be used self-consistently. The dimensionless quantity on the right side of (\ref{ZETA}) is therefore ``large'' in this sense. Whether the decay of an extremal AdS$_4$-Kerr-Newman black hole can lead to the emission of a black hole satisfying (\ref{ZETA}) depends on the details of the decay process, which are currently not well-understood.

In any case, it is clearly very likely that classical Cosmic Censorship will be violated by \emph{some} emitted black holes, and therefore one will have to consider how to modify Censorship in order to accommodate this in such cases. In doing so, one will have to take into account the presence of ``angular horizons,'' which may well persist even in the quantum-corrected case.

\addtocounter{section}{1}
\section* {\large{\textsf{8. Conclusion }}}
Consider an extremal AdS$_4$-Kerr-Newman black hole which satisfies the following two conditions:
\vskip0.2truecm
[1] Its angular momentum to mass ratio $\mathcal{A}$ satisfies $\mathcal{A}/L < 1$.
\vskip0.2truecm
[2] It is stable against emission of superradiant modes.
\vskip0.2truecm
We have found that if such a black hole emits a smaller, neutral rotating black hole along its axis of symmetry, then the angular momentum to mass ratio of the emitted black hole (which will itself be \emph{stable} against a superradiant instability) must satisfy $a/L > 1.$ That is, there is a sort of \emph{reciprocity} between the (specific) angular momenta of the two black holes; this is literally correct in the limit of large charge.

The existence of this reciprocity depends crucially on condition [2]. The fact that reciprocity takes such a simple form is a strong hint that there is some kind of relationship between stability against superradiance and the WGC. The nature of this relationship remains to be understood.

It is possible to escape these conclusions in two ways. First, the decay of the original black hole could be exclusively by emission of \emph{particles}, not black holes, with specific angular momenta greater than $L$; there is no particular reason to expect that such particles do not exist. The question, however, would be how the original black hole is able to distinguish black holes from such particles, and indeed some authors \cite{kn:kats} deny that any such distinction exists.

The second possibility is that the decay is \emph{only} possible by means of emission of charged, non-rotating objects. Apart from its intrinsic implausibility, there are technical reasons for doubting that this is possible. Examining equations (\ref{G}) and (\ref{GAG}), we see that, if $a/L$ is close to unity, then the minimal mass permitted by Cosmic Censorship is extremely sensitive to the angular momentum of the black hole, but correspondingly insensitive to the charge (because $\mathcal{P}$ always occurs in combination with a factor of $|1 - [a/L]|$). It is hard to believe that, if an extremal black hole with $a/L$ close to unity is obliged to decay, it will never do so in the most effective way, by losing a small quantity of angular momentum, rather than a very large quantity of charge. If it is so, then the actual mechanism of ``WGC decay'' must be very unusual.

Leaving these arguments to one side, we see that the ``black hole'' version of the Weak Gravity gives a concrete mechanism for producing exotic AdS black holes with ``angular horizons''. String-theoretic arguments suggest that the values of $a/L$ for these black holes cluster around $1.225$. The angular horizon for such a black hole is located at colatitude about 35 degrees away from both poles; the region enclosed by the angular horizon is far from the equator, but it is still substantial, and not likely to be banished altogether by a more complete quantum-gravitational treatment. The properties of these strange objects merit further attention.

\addtocounter{section}{1}
\section*{\large{\textsf{Acknowledgement}}}
The author is grateful to Dr. Soon Wanmei for useful discussions.

\end{document}